\documentclass[9pt,shortpaper,twoside,web]{ieeecolor}
\usepackage{generic}
\usepackage{cite}
\usepackage{amsmath,amssymb,amsfonts}
\usepackage{algorithmic}
\usepackage{graphicx}
\usepackage{textcomp}
\usepackage[colorlinks=true,linkcolor=blue]{hyperref}%
\usepackage{multirow}
\usepackage{color}
\newcommand{\tabincell}[2]{\begin{tabular}{@{}#1@{}}#2\end{tabular}}  
\usepackage[switch]{lineno}

\def\BibTeX{{\rm B\kern-.05em{\sc i\kern-.025em b}\kern-.08em
    T\kern-.1667em\lower.7ex\hbox{E}\kern-.125emX}}
\begin{document}
\title{HDL: Hybrid Deep Learning for the Synthesis of Myocardial Velocity Maps in Digital Twins for Cardiac Analysis}
\author{Xiaodan Xing, Javier Del Ser,~\IEEEmembership{Senior Member, IEEE}, Yinzhe Wu, Yang Li, Jun Xia, Lei Xu, David Firmin, Peter Gatehouse, and Guang Yang,~\IEEEmembership{Senior Member, IEEE}
\thanks{This work was supported in part by the European Research Council Innovative Medicines Initiative [DRAGON, H2020-JTI-IMI2 101005122], in part by the AI for Health Imaging Award [CHAIMELEON, H2020-SC1-FA-DTS-2019-1 952172], in part by the British Heart Foundation [Project Number: TG/18/5/34111, PG/16/78/32402], in part by the MRC [MC/PC/21013], and in part by the UK Research and Innovation Future Leaders Fellowship [MR/V023799/1]. J. Del Ser was supported by the Department of Education of the Basque Government for its funding support through the consolidated group MATHMODE [IT1294-19]. J. Xia was supported by the Project of Shenzhen International Cooperation Foundation (GJHZ20180926165402083).}
\thanks{X. Xing, Y. Wu, D. Firmin, P. Gatehouse, and G. Yang are with the National Heart and Lung Institute, Imperial College London, UK (send correspondence to x.xing@imperial.ac.uk and g.yang@imperial.ac.uk).}
\thanks{P. Gatehouse and G. Yang are also with the Royal Brompton Hospital.}
\thanks{Javier Del Ser is with the Department of Communications Engineering, University of the Basque Country UPV/EHU, 48013 Bilbao, Spain and TECNALIA, Basque Research and Technology Alliance (BRTA), 48160 Derio, Spain.}
\thanks{Y. Li is with the School of Automation Sciences and Electrical Engineering, Beihang University, Beijing, China.}
\thanks{J. Xia is with the Department of Radiology, Shenzhen Second People’s Hospital, Shenzhen, China.}
\thanks{L. Xu is with the Department of Radiology, Beijing Anzhen Hospital, Capital Medical University, Beijing, China.}}

\maketitle

\begin{abstract}
Synthetic digital twins based on medical data accelerate the acquisition, labelling and decision making procedure in digital healthcare. A core part of digital healthcare twins is model-based data synthesis, which permits  the generation of realistic medical signals without requiring to cope with the modelling complexity of anatomical and biochemical phenomena producing them in reality. Unfortunately, algorithms for cardiac data synthesis have been so far scarcely studied in the literature. An important imaging modality in the cardiac examination is three-directional CINE multi-slice myocardial velocity mapping (3Dir MVM), which provides a quantitative assessment of cardiac motion in three orthogonal directions of the left ventricle. The long acquisition time and complex acquisition produce make it more urgent to produce synthetic digital twins of this imaging modality. In this study, we propose a hybrid deep learning (HDL) network, especially for synthetic 3Dir MVM data. Our algorithm is featured by a hybrid UNet and a Generative Adversarial Network with a foreground-background generation scheme. The experimental results show that from temporally down-sampled magnitude CINE images (six times), our proposed algorithm can still successfully synthesise high temporal resolution 3Dir MVM CMR data (PSNR=42.32) with precise left ventricle segmentation (DICE=0.92). These performance scores indicate that our proposed HDL algorithm can be implemented in real-world digital twins for myocardial velocity mapping data simulation. To the best of our knowledge, this work is the first one in the literature investigating digital twins of the 3Dir MVM CMR, which has shown great potential for improving the efficiency of clinical studies via synthesised cardiac data.
\end{abstract}


\section{Introduction}
\label{sec:introduction}
Synthetic digital twins for medical data model the human body and provide realistic monitoring and evaluation data without invasion towards real patients. For a digital twin to faithfully model the complexity of the human body, data-based techniques are needed to learn and synthesise the vital signals that stem from its different parts. These techniques, which are crucial for actionable digital twins for healthcare, can be done by using complex numerical simulations or, instead, learned from real data by means of generative modelling. Methods such as UNet, Generative Adversarial Networks (GAN) \cite{beaulieu2019privacy} and their variations \cite{ge2021x,thermos2021controllable,xu2021synthesis} can provide synthetic medical data. However, the literature is scarce in algorithms for cardiac data synthesis. 

\begin{figure*}[!t]
\centerline{\includegraphics[scale=0.35]{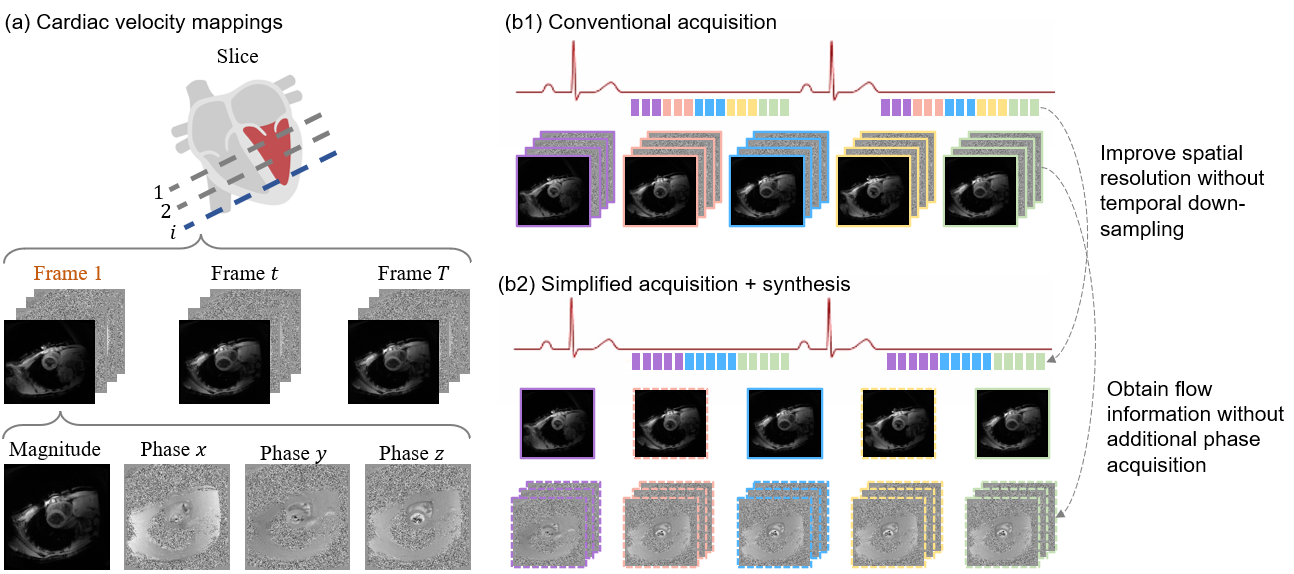}}
\caption{(a) An example of cardiac velocity mappings and (b) how synthesis algorithms help velocity mapping acquisition. Here in panel (a), for one slice in the left ventricle, $T$ frames were reconstructed per cardiac cycle. Each frame ($t=1,2,..,T$) in the slice is composed of one magnitude image and three phase images. The $x$ and $y$ directions are the in-plane directions while $z$ is the through-plane direction. In panel (b1), we represented the k-space sampling lines as colourful blocks and the velocity mapping images at each time point are reconstructed from k-space sampling with corresponding colours during two heartbeats (R-R intervals). More colourful blocks, i.e., more k-space sampling, produces images with a high spatial resolution. As panel (b2) shows, by synthesising intermediate magnitude images and all phase images, we can accelerate and simplify the traditional acquisition procedure. We hypothesise that the synthesised velocity mapping images can produce accurate velocity assessment results.}
\label{fig1}
\end{figure*}

Three-directional CINE multi-slice myocardium velocity mapping (3Dir MVM) is a cardiac MR (CMR) technique providing both magnitude and phase information of myocardium movements. 3Dir MVM is \textbf{not} the routinely acquired flow CMR, but provides the myocardium movement information. Patterns of myocardium movements provide important information for cardiac disease diagnosis. A 3Dir MVM usually acquires a series of MR images through one cardiac cycle. As is shown in Fig. \ref{fig1}, each time point at the image series is named as a frame, and each frame is usually composed of four MR images: one magnitude image and three phase images of three orthogonal directions. Pixels of these phase images are linearly related to their velocity by a constant called velocity encoding, i.e., $v_{enc}$ \cite{RN4}. By analysing the spatial information in magnitude data and velocity information in the phase data, researchers could localise and quantify myocardial motion patterns \cite{RN28}, as well as to measure myocardial functions. For example, myocardial motion abnormalities can happen due to aortic stenosis \cite{RN29} and other cardiovascular diseases.

However, the applications of 3Dir MVMs are limited because accurate and reproducible assessments may require images with high temporal resolution and precise and objective left ventricle (LV) segmentation. Acquiring these high-resolution images and segmentations requires massive consumption of time and human labour. Besides, data sharing and analysing on real patients might increase the risk of privacy information leakage. Thus, synthesising realistic vital information of 3Dir MVM data is of vital importance in cardiac motion analysis and digital twins. Deep learning-based image synthesis is a popular method of producing such signals, becoming nowadays a research area of utmost relevance for the creation of digital twins for cardiac healthcare.

There are few algorithms on 3Dir MVM digital twin synthesis, even though these algorithms are clinically important. Three major difficulties caused this. The first difficulty is in improving the temporal resolution of digital twins. To obtain cardiac motion patterns accurately, the temporal resolution must be high enough (at least 9-13 frames per cardiac cycle) \cite{RN5}. The second difficulty exists in ROI labelling. Since we do have ROI labelling in real-world datasets, we need to investigate how to transfer this labelling information to our synthetic datasets. The third difficulty is in phase image synthesis because the values of pixels on phase images are vital for motion analysis, whereas there are many noisy pixels in phase images.

To address the aforementioned problems, we propose a novel hybrid deep learning (HDL) framework to synthesise digital 3Dir MVM CMR images with a high temporal resolution, together with accurate segmentation results. The inputs of the HDL model are magnitude CINE images with a low temporal resolution (the images with solid borders in Fig. \ref{fig1} (b2)), and the outputs of HDL models are interpolated magnitude images and synthetic phase images (the images with dashed borders in Fig. \ref{fig1} (b2)). The inputs and outputs images together serve as synthetic digital twins of 3Dir MVM data and produce myocardium velocity assessment results.  

To summarise, the novelty of this work is three-fold: 1) We introduced an HDL model to synthesise 3Dir MVMs with a high temporal resolution using magnitude data only; 2) Our work is able to produce segmentation masks with high accuracies; and 3) We conducted myocardial velocity assessment from synthetic 3Dir MVMs and proved the effectiveness of synthetic 3Dir MVM data. To the best of our knowledge, our work is the first study on synthetic 3Dir MVM in the literature. Our work brings a new solution towards the limitation of 3Dir MVMs acquisitions.

The paper is organised as follows: Section \ref{sec:relatedw} introduces some related works regarding the synthesis of medical images; Section \ref{sec:method} details the methodology of our proposed HDL; Section \ref{sec:experiments} presents the data and parameter settings of our experiments, as well as the experimental results; discussions on parameter and model selections are in Section \ref{sec:discussion}; Section \ref{sec:conclusion} concludes our study. 
\section{Related Works}
\label{sec:relatedw}
The first problem in real-world 3Dir MVM acquisition is the trade-off between temporal and spatial resolution. Due to the reconstruction nature of MRI acquisition, an inevitable trade-off exists between the temporal and spatial resolution for CINE data acquisitions. As shown in Fig. \ref{fig1} (b1), during one R-R interval if more time points are acquired, fewer lines can be sampled in the k-space, leading to a lower spatial resolution. It is worth noting that in the literature, time point is also addressed as phase, while to avoid misunderstanding and to distinguish this with phase data, we will call it time points in the following sections. Although the view-sharing technique \cite{RN24} could produce additional time points during one R-R interval, it introduced motion artefacts \cite{RN25}. Therefore, how to improve the temporal resolution of velocity mappings without compromising the spatial resolution is still an open question to be investigated. 

A digital twin of CINE images series with a higher spatial resolution can be synthesised by frame interpolation algorithms. Frame interpolation algorithms can improve the temporal resolution of image series by predicting the intermediate frames from existing frames. These algorithms include conventional methods such as linear interpolation and optical flow \cite{RN16}, phase constraints \cite{RN17} and deep learning based methods, most of which are UNet based models, with optical flow estimation \cite{RN18} or with direct interpolation prediction \cite{RN30}. However, these methods are designed for natural image processing, where the moving objects in these tasks are rigid, while the myocardium moves elastically.

Despite the acquisition time, complex scanning protocols of 3Dir MVM CMR also limit the applications of this technology. Moreover, parameter selections for 3Dir MVM CMR acquisition can be complex. For example, the encoding velocity $v_{enc}$ should match the real peak velocity. Higher $v_{enc}$ may cause more noises \cite{RN6} while if the encoding velocity is lower than the true peak velocity, aliasing artefacts may occur \cite{RN4}.   

We can also avoid scanning parameter selections by synthesising digital twins of phase images without acquiring them directly. Considering that magnitude images are relatively easier to acquire, and many public datasets are containing only magnitude images (facilitating for further segmentation model training) \cite{RN23}, we choose magnitude images as our source images and phase images as our target images to be synthesised. Cross-modality image synthesis algorithms can provide a solution for this task, e.g., UNet \cite{RN21} and Generative Adversarial Networks (GAN) \cite{RN19}\cite{RN20} showed promising results in cross-modality synthesis. However, these methods are not validated on 3Dir MVM data either. 
\section{Materials and Methods}
\label{sec:method}
\begin{figure*}[!t]
\centerline{\includegraphics[scale=0.4]{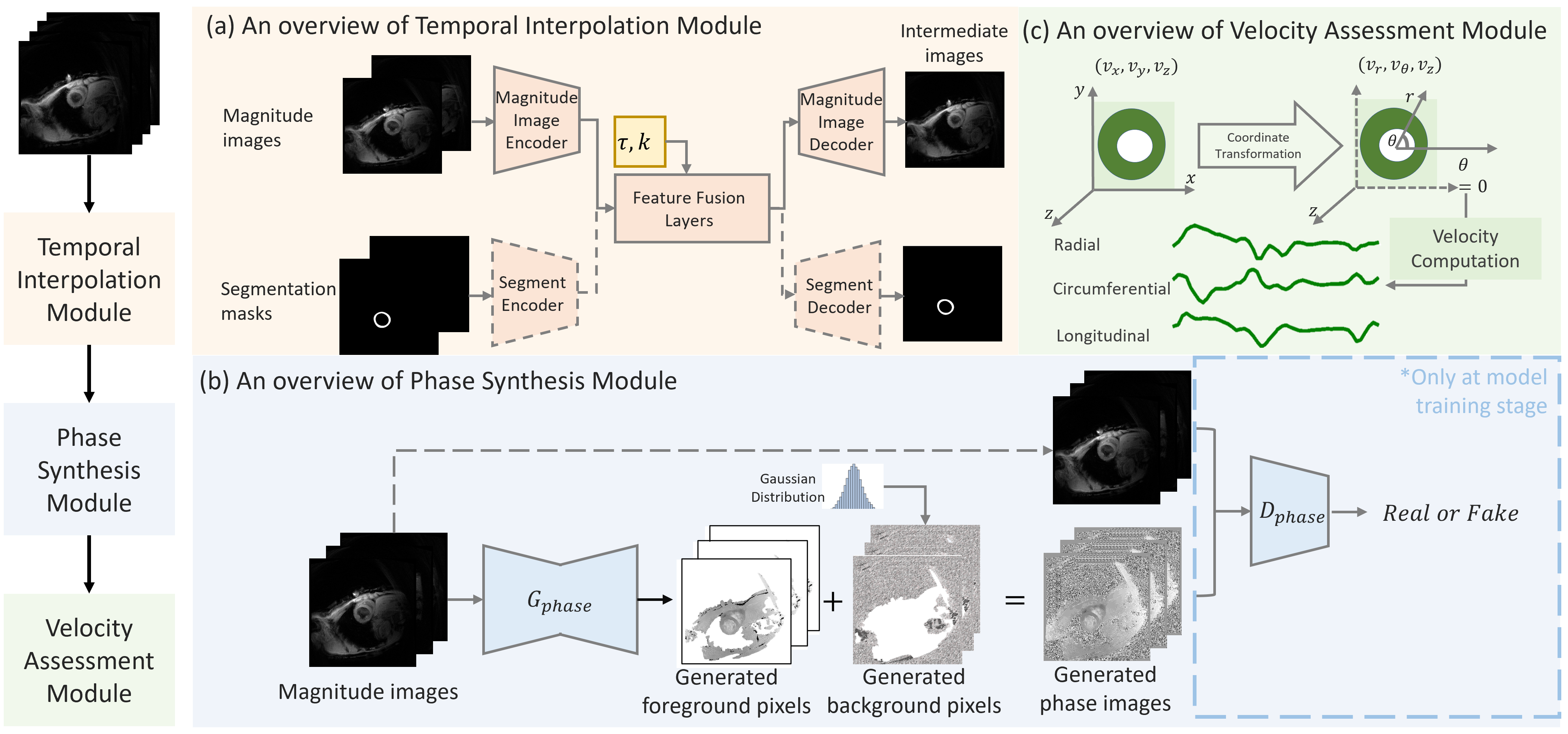}}
\caption{A schematic overview of our proposed hybrid deep learning digital twin framework. Our framework is composed of three modules: 1) Temporal Interpolation Module: an UNet based model to improve the temporal resolution by interpolating intermediate magnitude images and generating corresponding segmentation results (if plausible); 2) Phase Synthesis Module: a pix2pix based generative adversarial network (GAN) to synthesise phase images from magnitude images; and 3) Velocity Assessment Module: a pipeline to assess the velocity and motion information from the synthesised velocity mapping data.  \textcolor{black}{Different modules in our method are performed sequentially. The network of each module is trained separately. During inference, our model receives a series of magnitude images and interpolates $K$ images from images at $t=\tau$ and $t=\tau+1$. Here $k\in\{0,1,2,...,K\}$. With the temporal up-sampled images, we compute the corresponding phase images and combine both magnitude and phase images as synthetic velocity mappings. Finally, with the synthetic velocity mappings and the velocity assessment module, we compute velocity curves in the left ventricle and present velocity assessment results.}}
\label{fig2}
\end{figure*}

An overview of the proposed three-step hybrid deep learning framework is shown in Fig. \ref{fig2}. In this section, we will introduce the details of these three modules, respectively.
\subsection{Temporal Interpolation Module}
\label{sec:interpmethod}
We use magnitude images and corresponding segmentation masks to interpolate intermediate magnitude images. Mathematically, we would like to interpolate magnitude images $\{\tilde{M}_{\tau+k}|k=1,2,...,K\}$ from images $M_{t=\tau}$ and $M_{t=\tau+K+1}$. In the following sections, magnitude images $M_{t=\tau}$ and $M_{t=\tau+K+1}$ are \emph{existing images}, and images $\{M_{\tau+k}|k=1,2,...,K\}$ are \emph{ground truth images to be interpolated}. Images $\{\tilde{M}_{\tau+k}|k=1,2,...,K\}$ are \emph{images interpolated}, which are also the outputs of our temporal interpolation module. Considering the periodicity of cardiac motions, when $\tau+K+1>50$ we used $M_{t=\tau+K+1}=M_{t=\tau+K+1-50}$ , and  $M_{t=\tau+K+1}=\tilde{M}_{t=\tau+K+1-50}$ during testing. Discarding $\{M_{\tau+k}|k=1,2,...,K\}$ is also to temporally down-sample magnitude image series $K$-times. 
\begin{figure}[h]
\centerline{\includegraphics[width=\columnwidth]{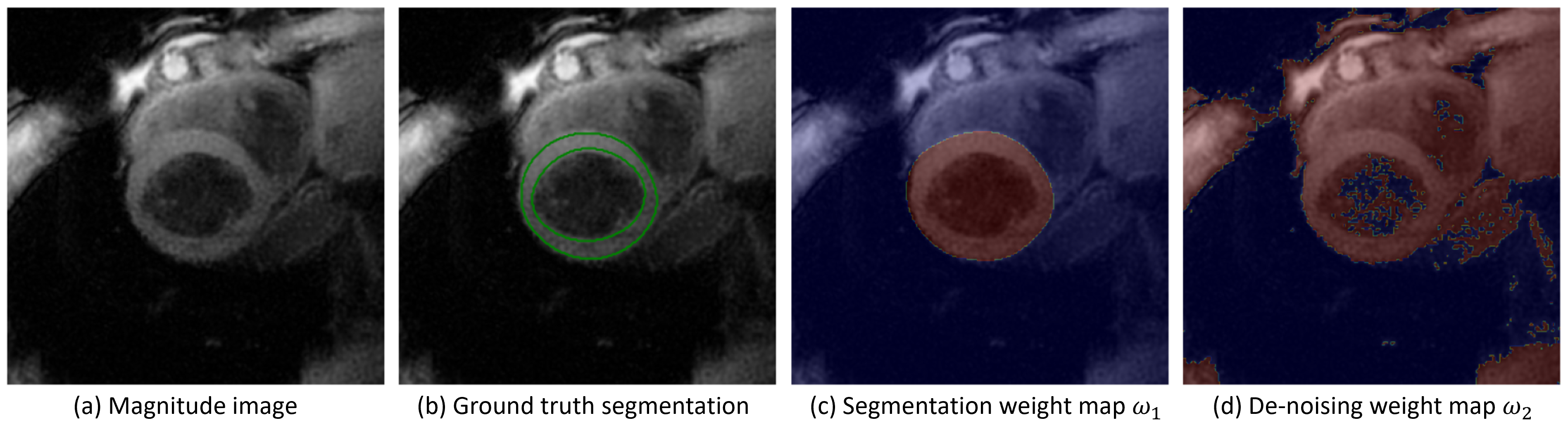}}
\caption{Weight maps for the MAE loss functions. Highlighted red pixels in weight maps have weights of 1, while other blue pixels have weights of 0.1. $\omega_2$ is a padded patch from the outer contour of ground truth segmentation. $\omega_2$ is computed from magnitude images (ranging from [-1,1]), from which pixels with values $>-0.95$ are highlighted.}
\label{fig3}
\end{figure}

The target of this temporal interpolation is a conditional synthesis task, with a condition of $\tau$ and $k$. It is because that cardiac cycles for all healthy subjects are similar; and $\tau$ and $k$ are decisive information when generating intermediate images. We up-sample $\tau$ and $k$ as feature maps and then concatenate these two maps together. This matrix is named temporal condition maps. 

We use a multi-head multi-tail R2UNet \cite{RN7}\cite{RN27} to interpolate temporal frames. The multi-head and multi-tail structure allows independent layers when encoding and decoding inputs from different domains. For example, magnitude images and segmentation masks are from different image domains; if encoded and decoded dependently using the same parameters, artefacts may occur (see Fig. \ref{fig6} in Section \ref{sec:interpexp}). The recurrent structure in R2UNet improves the efficiency of network parameters as well as the segmentation performance. 

Despite temporal condition maps and multi-head multi-tail structure, we adjust the loss functions by adding weights on the mean absolute error (MAE) loss function. We use a weighted MAE with two masks, as is shown in Fig. \ref{fig3}. The weight map $\omega_1$ highlights the region inside the left ventricular myocardium, which is our ROI in the following velocity assessment. The weight map $\omega_2$ highlights non-background pixels in the magnitude images. Mathematically, our weighted-MAE function is defined as below
\begin{equation}
L(\tilde{M},M)=\mathrm{Mean}(\omega_1\times||\tilde{M}-M||) + \mathrm{Mean}(\omega_2\times||\tilde{M}-M||).
\end{equation}
Besides, the masked MAE will be addressed as MAEW1 and MAEW2 in the following sections.

It is of note that unlike the phase generator in Section \ref{sec:phasemethod}, which is trained with a GAN based architecture, we discard the discriminator loss during temporal interpolation. This is because phase generation is a cross-modality synthesis task, and the discriminator loss is designed to approximate the intensity distribution of phase images. However, in the temporal interpolation, the intensity distribution of all magnitude images is directly encoded in inputs and our model shall only focus on the correct positioning of each tissue. In this situation, we assume that MAE loss is more straightforward to be optimised compared to the GAN loss. In the discussion section, this assumption is validated using a comparison experiment. 
\subsection{Phase Image Synthesis Module}
\label{sec:phasemethod}
As we noticed that the velocities, although not directly shown, are encoded in the moving patterns of the magnitude images. According to the spatial movements and the depth changing of pixels, we assume that the in-plane and through-plane velocities can be synthesised, and the directions of the velocities and the exact values of the phase images can be then defined. Based on these hypotheses, we use a pix2pix \cite{RN11} model to synthesise phase images from magnitude images. We use an R2UNet as the generator and a PatchGAN \cite{RN8} discriminator with four convolutional blocks as the discriminator.
The network input contains not only the corresponding magnitude image for the phase image synthesis but also contains two neighbouring magnitude images in chronological order because phase images are computed from the pattern of movements. For example, to synthesise the phase image $P_{t=1}$, the inputs to phase generation module are magnitude images $M_{t=0}$, $M_{t=1}$ and $M_{t=2}$; considering the periodicity of cardiac motions, to synthesise the phase image $P_{t=T}$, the inputs to phase generation module are magnitude images $M_{t=T-1}$, $M_{t=T}$ and $M_{t=0}$; for magnitude series which are temporally down sampled, the input magnitude images are $\tilde{M}_{t=0}$, $\tilde{M}_{t=1}$ and $\tilde{M}_{t=2}$. 
\begin{figure}[h]
\centerline{\includegraphics[width=\columnwidth]{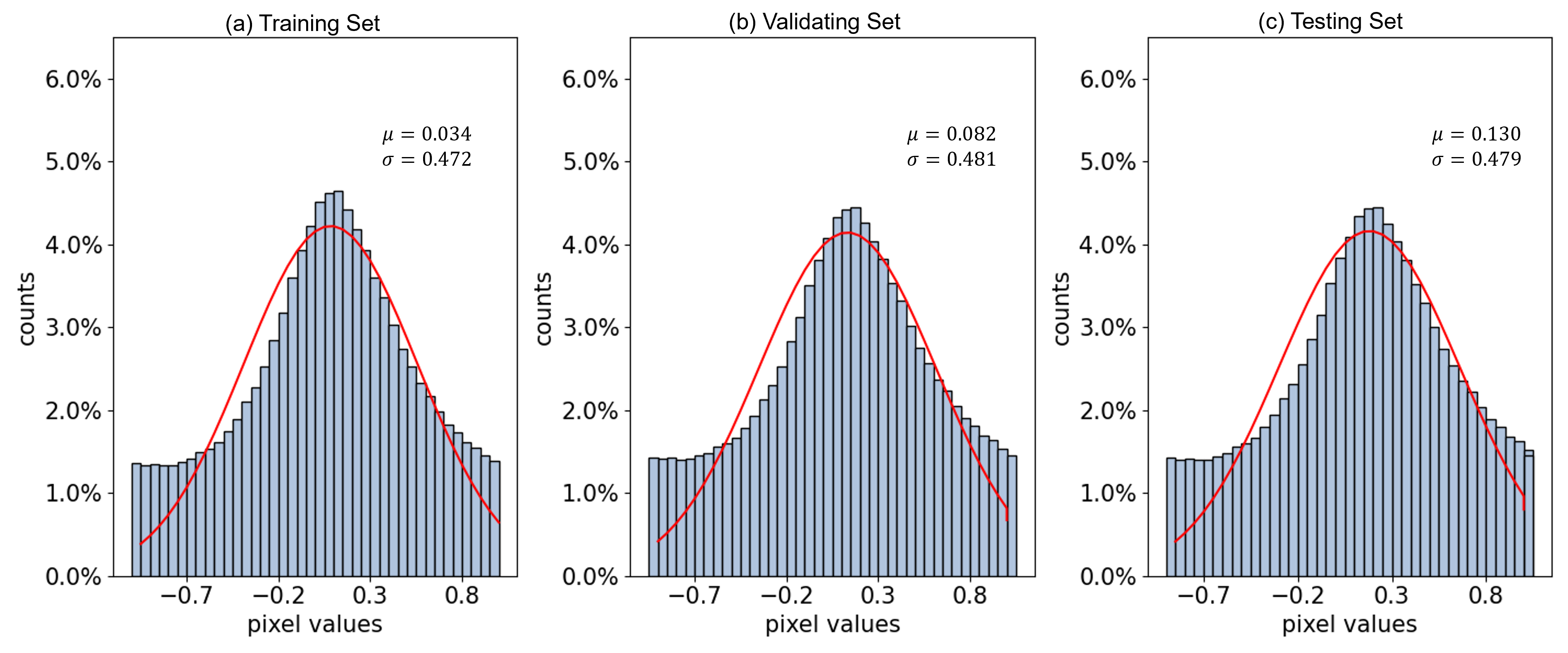}}
\caption{The histograms (blue bins) and approximated Gaussian distributions (red curves) in normalised training, validating, and testing datasets. Original pixel values of phase images are between [0,4096], while we normalised them between a range of [-1,1]. In our experiment, we used the approximated distribution in the training set ($\mu=0.034$,$\sigma=0.034$) to randomly generate background noises.}
\label{fig4}
\end{figure}
We propose a foreground-background generation scheme for phase image generation. A greyscale phase image is typically displayed with three parts: irregular white and black pixels which are the background noises, uniform grey pixels which are stationary tissues and uniform white and black pixels which are the velocity in two directions. Apparently, distance measurements such as Mean Square Error (MSE) are not suitable for background noise synthesis. Thus, we use different generators for tissue pixels and background noise pixels. For stationary and moving tissue pixels, which are foreground pixels, we use the deep learning generator. For background noise pixels, we sample from a Gaussian distribution. The background noise Gaussian is shown in Fig. \ref{fig4} (a). We also plotted the noise distributions in the validating and testing set in Fig. \ref{fig4} (b) and (c) to make sure the Gaussian distribution is generalisable.

We use magnitude images to split foreground pixels and background pixels. Background pixels in magnitude images are shown as black pixels (pixel value $<-0.95$ with all values in [-1,1]). As is shown in Fig. \ref{fig3} (d), foreground and background pixels are easy to demarcate with simple thresholding.  
\subsection{Velocity Assessment Module}
\label{sec:velocitymethod}
We use several velocity assessments scales to validate the performance of our synthetic 3Dir MVM. The pixel values of phase images are linearly correlated with the myocardial velocities, and for synthetic phase images, we re-normalized them into [0,4096] to avoid value shifting. Considering the circular shape of LV, we compute the velocity and then transfer velocities from the Cartesian coordinate system into a three-directional cylindrical coordinate system, as is shown in Fig. \ref{fig2} (c). Then we plot the curves of global radial ($v_r$), circumferential ($v_c$) and longitudinal ($v_z$) LV myocardial velocities. From the velocity curves, we analyse the peak systolic velocity (PSV), time to peak systolic velocity (TPSV), peak diastolic velocity (PDV), time to peak diastolic velocity (TPDV), and mean velocity (Mean) during one cardiac cycle \cite{RN9}. These scales are important markers for cardiac motion analysis in clinical settings \cite{RN14}. 
\section{Experimental Settings and Results}
\label{sec:experiments}
\subsection{Data}
We used an in-house 3Dir MVM dataset to train and validate our HDL model. Our in-house dataset was obtained from Royal Brompton Hospital, UK, including 19 healthy volunteers (mean age 38 years, range 24–59) \cite{simpson2014spiral}. 11 out of 19 subjects were acquired twice, giving us a final dataset with a size of 30.

The retrogated spiral phase velocity mapping sequence was implemented on a clinical scanner (3T, MAGNETOM VIDA, Siemens AG Healthcare Sector, Germany) equipped with an anterior cardiac 18-element matrix coil and a 32 element spine array. Typically, 12 elements of the cardiac coil and 8 of the spine coil were used for the in vivo acquisitions. Localised second-order shimming and frequency adjustment based on the signal from a user-defined adjustment box situated over the whole heart was performed to reduce off-resonance effects. 

The acquisition parameters are TR = 24 ms, TE = 4.4 ms and the FOVs of images are partly 360 mm and partly 400 mm. FA is 15 degrees by water selective excitation, and SLT = 8mm. The reconstructed TR is the average R-R interval during the 13-cycle BH, divided by 50 interpolated output frames. The acquired resolution is 1.41-1.56 mm, spatially interpolated by zero-filled FFT after regridding, to 0.70-0.78mm for output images. Velocity-encoding $v_{enc}$ is 30cm/s through-slice and 20cm/s for both in-plane axes. The CG-SENSE acceleration factor is 3 with Gadgetron reconstruction.

We selected 10 subjects (15 scans, 70 slices) for training, 5 (10 scans, 48 slices) for validating and 4 (5 scans, 24 slices) for testing. The study was approved by the local ethics committee, and all patients provided written informed consent. The ground truth segmentation of the LV was performed and validated by a qualified CMR physicist. 
\subsection{Experimental Parameter Settings}
\noindent\textbf{Temporal Interpolation Module.} We used a multi-head multi-tail R2UNet structure for temporal data synthesis. During the training stage, we used an Adam optimiser with a learning rate of 0.001 and a batch size of 32. To accelerate the loss convergence, a skip connection was added between the input layer and the output layer of the R2UNet model. Since we have the ROI segmentation masks for all subjects, we cropped the image around ROIs from $512\times512$ to $256\times256$ and discarded non-ROI regions. We trained our temporal generator network with real phase images and selected the model that performed best on the validating set with synthetic images. The independent testing set is then used for the following velocity assessment. 

We presented the comparison results among four methods including linear interpolation, optical flow interpolation, R2UNet and our method. Considering that the ground truth image to be interpolated is $ M_{t=\tau+k}$, and its corresponding segmentation mask is $S_{t=\tau+k}$, the input existing images and segmentation masks are $M_{t=\tau}$, $M_{t=\tau+K+1}$, $S_{t=\tau}$, and $S_{t=\tau+K+1}$. For linear interpolation, the interpolated result is then
\begin{equation}
\begin{aligned}
&\tilde{M}_{t=\tau+k}=\frac{k}{K+1}M_{t=\tau}+(1-\frac{k}{K+1})M_{t=\tau+K+1},\\
&\tilde{S}_{t=\tau+k}=\frac{k}{K+1}S_{t=\tau}+(1-\frac{k}{K+1})S_{t=\tau+K+1}.
\end{aligned}
\end{equation}	

For the optical flow method, we used the Horn-Schunck \cite{RN16} algorithm. The segmentation masks of the optical flow method are generated from the same optical flow computed from the magnitude images. For the R2UNet model, we used the same set of training parameters as the proposed model. The input of the R2UNet has 6 channels, including $M_{t=\tau}$, $M_{t=\tau+K+1}$, $S_{t=\tau}$, and $S_{t=\tau+K+1}$, and the two temporal condition maps. The output of the R2UNet is a two-channel map including both the synthesised images and the segmentation results.
\begin{table*}\centering
\caption{Performance comparison of the synthesised phase data results.}
\label{tab2}
\begin{tabular}{ccccccc}
\hline
 & & 		MSE$\times$100(Std$\times$100)&	MSEW1$\times$10(Std$\times$10)&	MSEW2$\times$100(Std$\times$100)&	PSNR(Std)&	SSIM(Std)\\
 \hline
\multirow{2}{*}{$K=0$}	&Vanilla Pix2pix	&4.08(0.53)	&0.29(0.10)	&0.19(0.16)	&42.02(0.61)	&0.26(0.02)\\
	&Proposed model	    &3.66(0.57)	&0.29(0.07)	&0.12(0.15)	&42.50(0.76)	&0.29(0.05)\\
\hline
\multirow{2}{*}{$K=3$}	&Vanilla Pix2pix	&3.90(0.53)	&0.29(0.10)	&0.24(0.16)	&42.22(0.60)	&0.25(0.02)\\

	&Proposed model	    &3.65(0.58)	&0.30(0.08)	&0.16(0.17)	&42.51(0.76)	&0.29(0.06)\\
\hline
\multirow{2}{*}{$K=6$}	&Vanilla Pix2pix	&4.10(0.43)	&0.31(0.16)	&0.51(0.09)	&42.00(0.51)	&0.25(0.26)\\

	&Proposed model	    &3.81(0.38)	&0.29(0.09)	&0.19(0.13)	&42.32(0.46)	&0.29(0.14)\\
\hline
\end{tabular}
\end{table*}

\noindent\textbf{Phase Generation Module.} The pix2pix model was trained with a learning rate of 0.002, a batch size of 12, and an input size of $256\times256$. Input magnitude and phase images are normalised between [-1,1]. We used Adam optimisers for the generator and the discriminator. We used a patch GAN based discriminator for our task and the patch size is $16\times16$.
\subsection{Temporal Interpolation Results}
\label{sec:interpexp}
\begin{figure}[h]
\centerline{\includegraphics[width=\columnwidth]{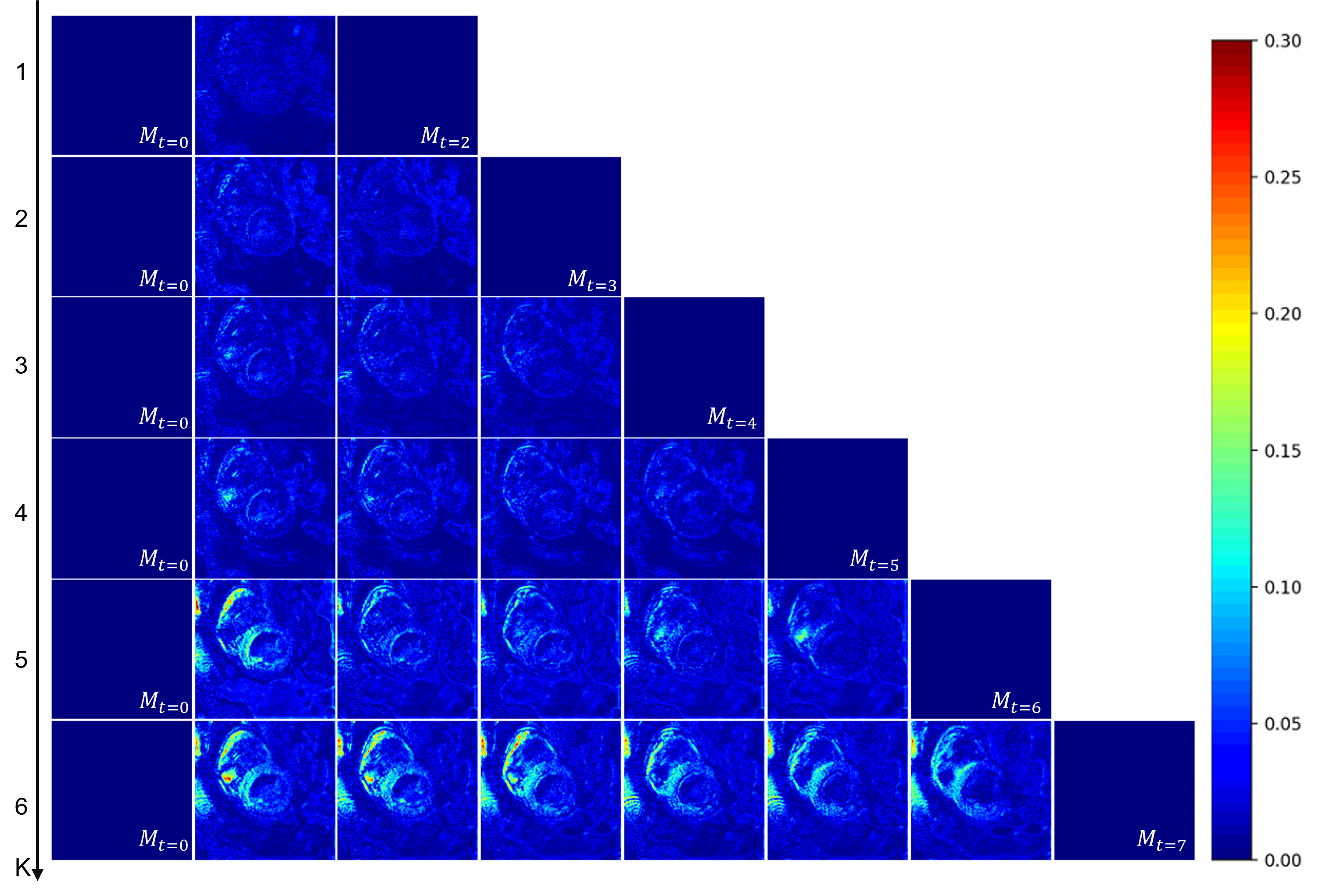}}
\caption{The error maps of the proposed method with $K=1,2,3,4,5,6$. Here we present the middle slice of an example subject, and we choose $\tau=0$, i.e., the first existing image is $M_{t=0}$. Images that existed after down-sampling are shown in a full blue patch.}
\label{fig5}
\end{figure}
As mentioned in Section \ref{sec:interpmethod}, our target is to interpolate magnitude images $\{\tilde{M}_{t=\tau+k}|k=1,2,...,K\}$ from $M_{t=\tau}$ and $M_{t=\tau+K+1}$ and generate corresponding segmentation masks simultaneously. We tested our temporal interpolation algorithm on $K=1,2,3,4,5 \mathrm{~and~} 6$. In Fig. \ref{fig5}, we presented the absolute error maps between the interpolated images and the ground truth images. By increasing $K$, we temporally down-sampled the magnitude image series with fewer existing images. At the first four columns, the errors of our interpolation mainly existed on background pixels (pixels not highlighted in Fig. \ref{fig3} (c) and (d)). However, when the down-sampling rate increased, more errors began to occur in ROI pixels.  
\begin{figure}[h]
\centerline{\includegraphics[width=\columnwidth]{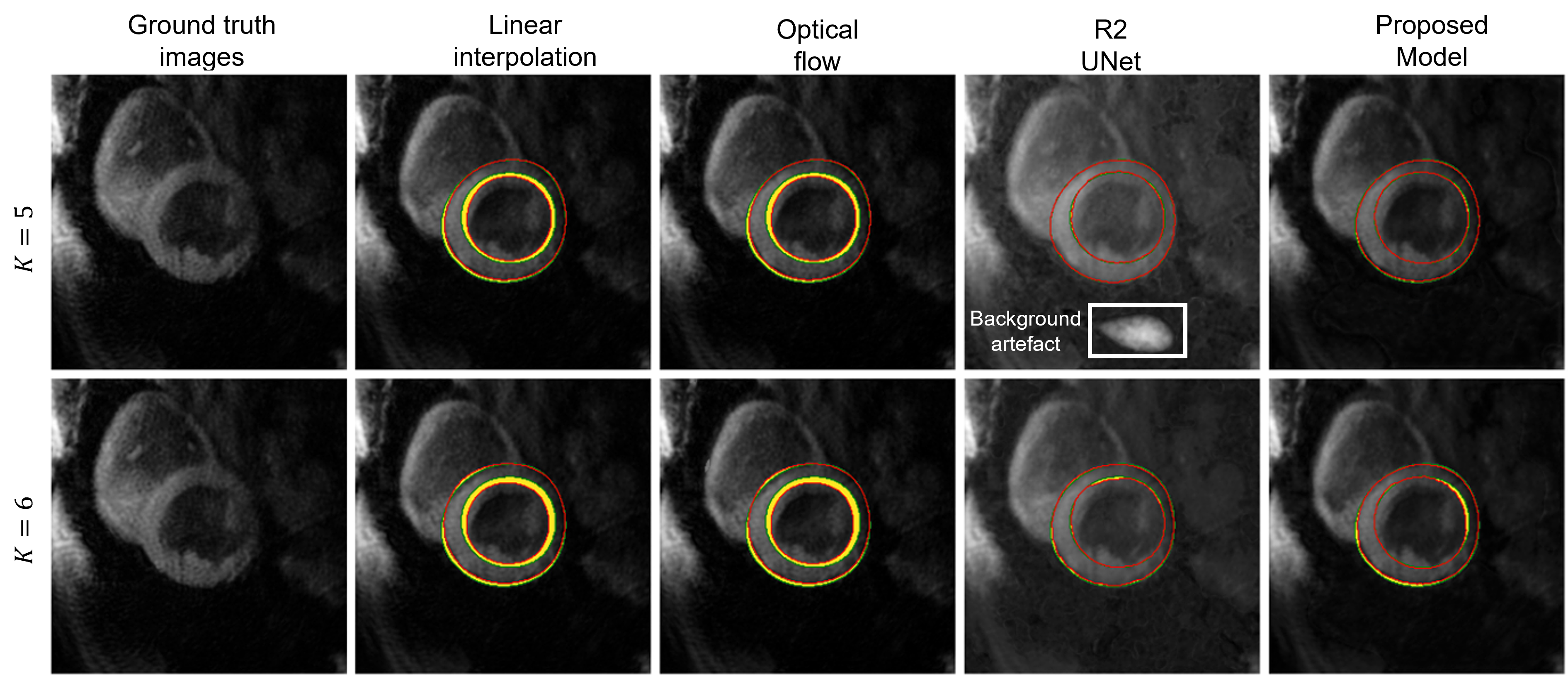}}
\caption{The LV segmentation results using different interpolation and synthesis methods. The image shown here is the MID slice of an example case and $t=1$. The red contours here represent the predicted segmentation, while the green contours represent the ground truth segmentations; the difference between these segmentations are highlighted in yellow; an artefact occurs in the R2UNet synthesised image because the R2UNet uses the same embedding parameters for inputs from multiple domains.}
\label{fig6}
\end{figure}
Compared to the R2UNet, our proposed temporal data synthesis module has independent encoders and decoders for magnitude images and segmentation masks. The independent encoders and decoders can facilitate the simultaneous generation of images from different domains. Fig. \ref{fig6} presents the synthesised magnitude images and corresponding segmentation results with $K=5$ and $K=6$, whose errors are more prominent than $K=1,2,3,4$. Although R2UNet did produce accurate segmentation results, the interpolated magnitude images possessed less image contrast compared to the ground truth images. This domain difference may be caused by the value differences between magnitude images and the segmentation masks. Moreover, the weighted loss function was not optimised on background pixels, which could induce artefacts (cropped in a white box in Fig. \ref{fig6}) in the background of R2UNet interpolated magnitude images. 
\subsection{Phase image generation results}
\label{sec:phaseexp}
\begin{figure}[h]
\centerline{\includegraphics[width=\columnwidth]{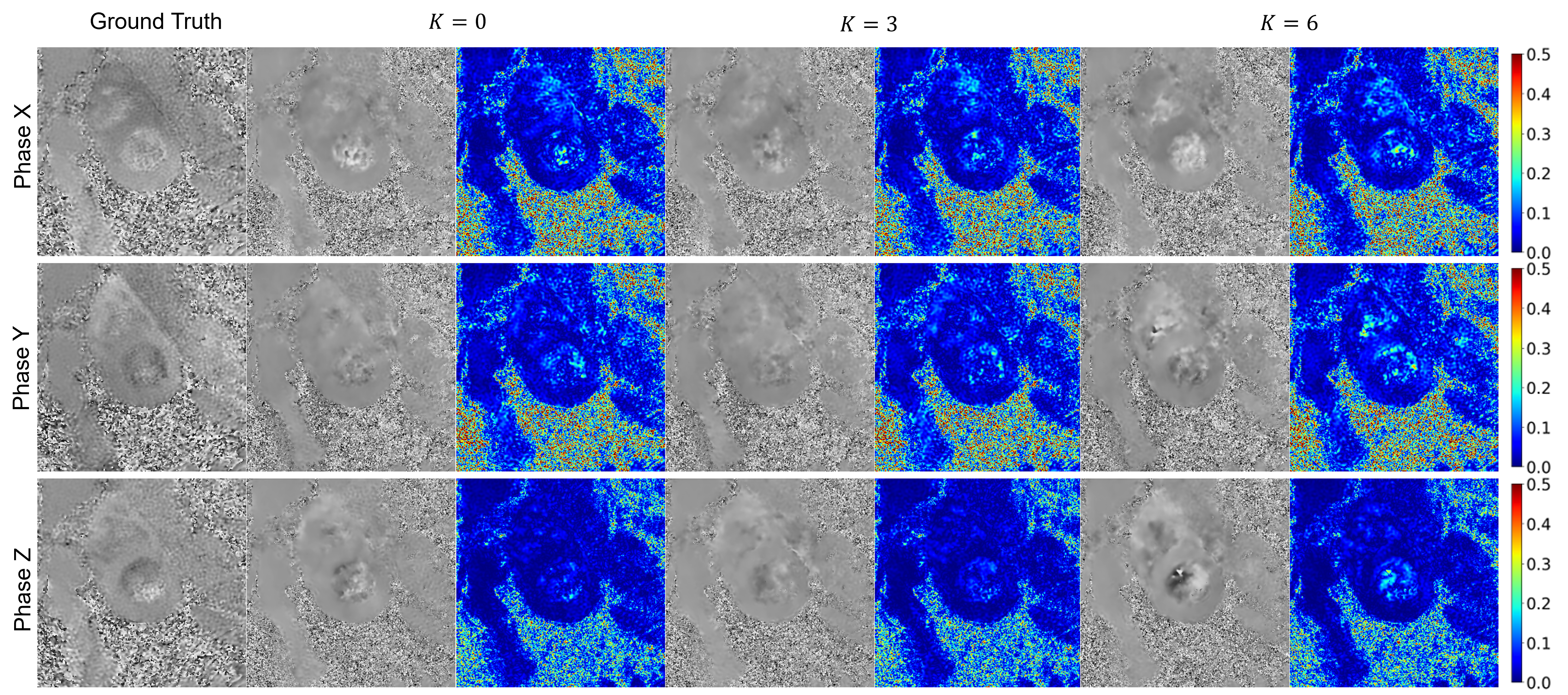}}
\caption{An example result of the proposed phase data synthesis module. The first column contains ground truth phase images, while other columns are the synthesised phase images with $K$-interpolated magnitude images and corresponding absolute difference maps. The images presented are from time point $t=1$ of an example subject.}
\label{fig7}
\end{figure}

We presented example results of our synthesised phase data in Fig. \ref{fig7}. As is shown, most errors occur on background pixels. As the temporal down-sampling factor $K$ increases, more errors occur on ROI regions.
To validate the weighted loss and foreground-background independent generation process, we compared the proposed method with vanilla pix2pix, where the noise pixels and tissue pixels of phase images are synthesised together from the generator. We compared the quality of synthetic phase images using three metrics including the mean squared error (MSE), the peak signal to noise ratio (PSNR), and the structural similarity index measure (SSIM). Table \ref{tab2} shows the mean MSE, PSNR and SSIM for all subjects in the testing dataset. 
\subsection{Velocity Assessment Results}
\label{sec:velocityexp}
\begin{figure}[h]
\centerline{\includegraphics[width=\columnwidth]{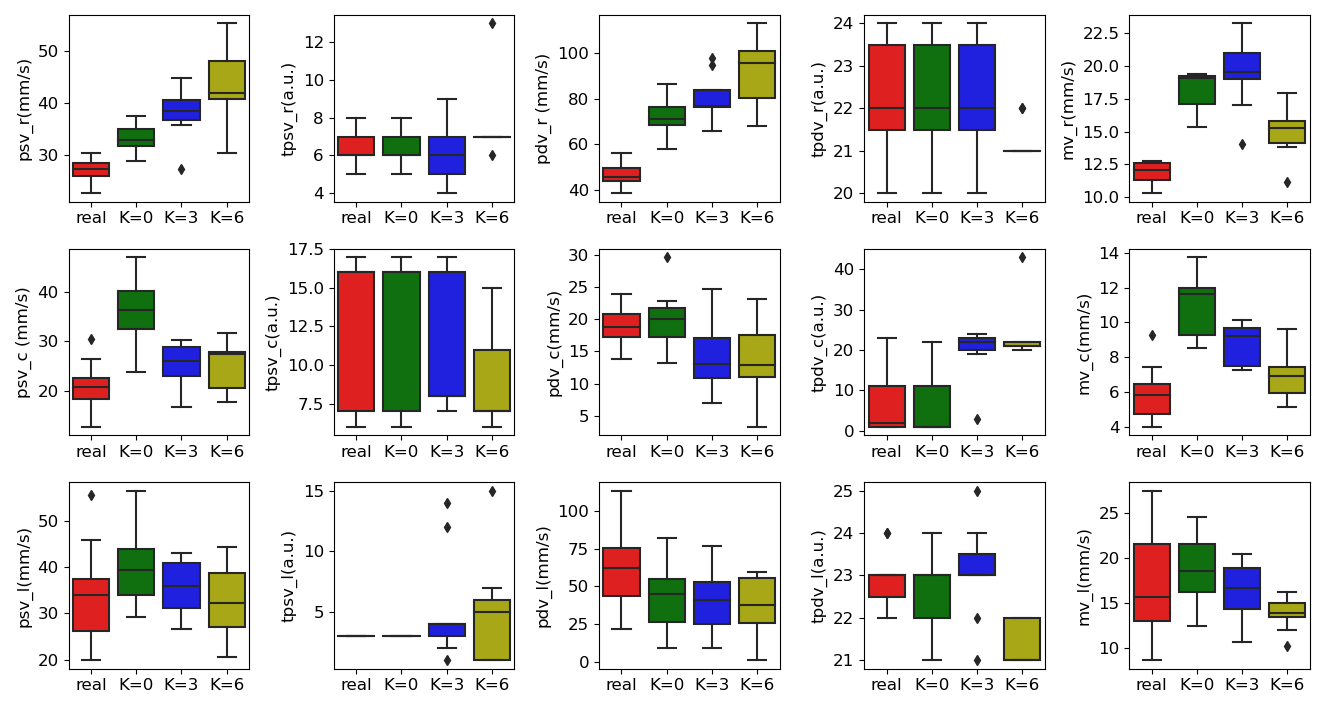}}
\caption{Box plots of velocity assessment results of real phase images (real), synthesised phase images from real magnitude images ($K=0$) and synthesised phase images from interpolated magnitude images ($K=3$ and $K=6$).}
\label{fig8}
\end{figure}

\begin{figure}[h]
\centerline{\includegraphics[width=\columnwidth]{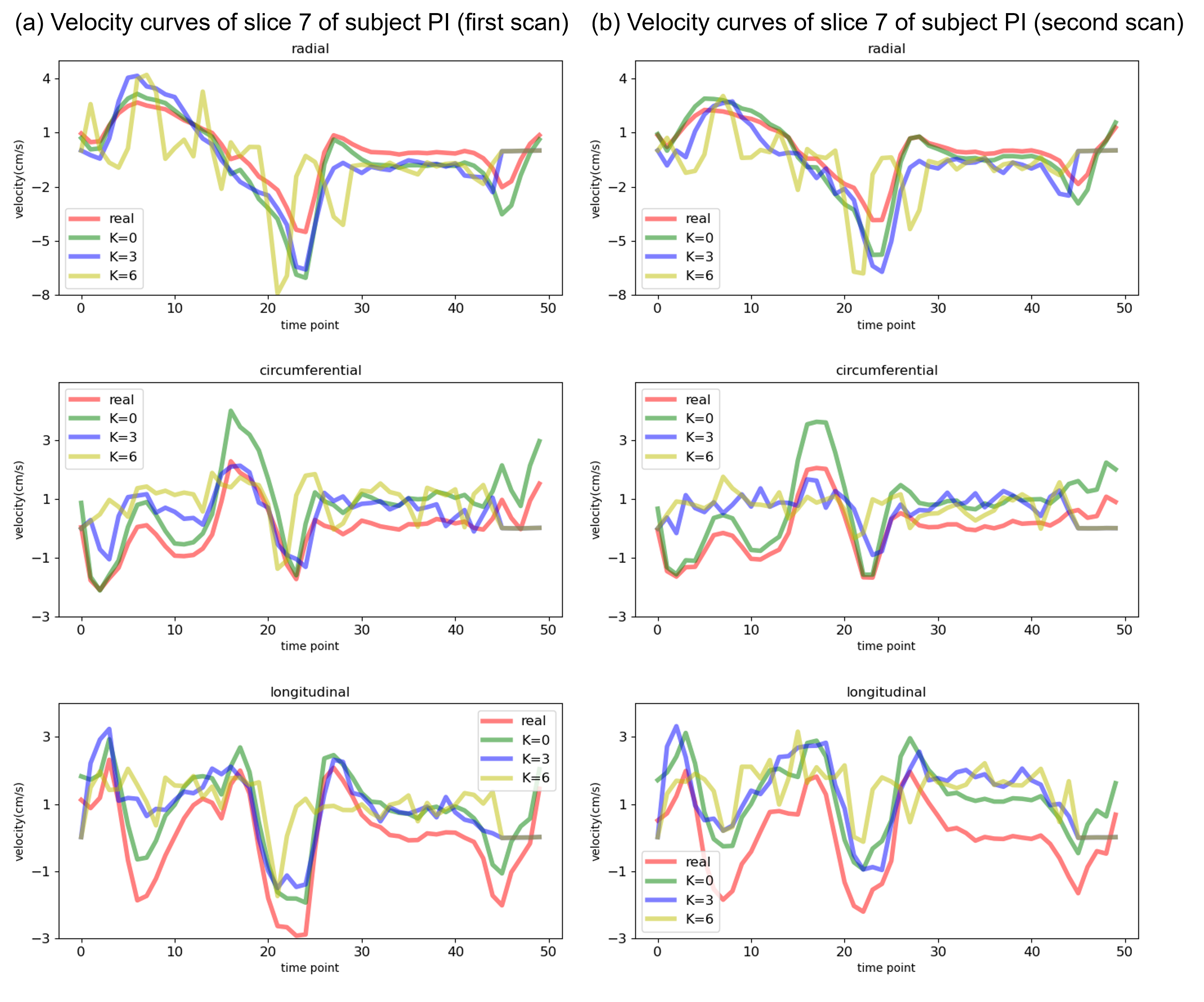}}
\caption{Velocity curves of the same subject PI from (a) the first scan and (b) the second scan. The y-axes of this figure represent values of the velocity, and the x-axes represent the normalised time point t. Since the time of one cardiac circulation varies between subjects, we normalised the physical time into 50 time points for a fair comparison.}
\label{fig9}
\end{figure}

Table \ref{tab3} shows the results of the velocity analyses on real phase images, synthesised phase images with real magnitude images as inputs and synthesised phase images with $K=3$ and $K=6$ interpolated magnitude images as inputs. We also plotted the velocity assessment results in Fig. \ref{fig8}. For radial PSV, radial PDV and radial mean velocity, velocity assessment results from the synthesised phase images are higher than real results. While the values and distributions for PSV, PDV and mean velocity in circumferential and longitudinal directions are similar. TPSV and TPDV indicate the time from $t=0$ to the peak velocities. Because the R-R interval for each subject is different, we normalized the physical time into 50 time points. The Pearson's correlations between predicted velocity curves and real curves are 0.99, 0.92, 0.48 separately with $K\in[0,3,6]$. In Fig. \ref{fig9}, we presented two example velocity curves from two scans of the same slice of an example subject. 
\begin{table}\centering
\caption{The velocity assessment results (Mean(Std)). Here, a.u. is arbitrary units and indicates the time points during one R-R interval.}
\label{tab3}
\begin{tabular}{cccccc}
\hline
 & & 		REAL&$K=0$&	$K=3$&	$K=6$\\
 \hline
\multirow{5}{*}{radial}&	\tabincell{c}{psv \\ mm/s}& 	\tabincell{c}{27.01\\(0.21)}& 	\tabincell{c}{33.12\\(0.23)}& 	\tabincell{c}{38.19\\(4.34)}& 	\tabincell{c}{44.03\\(6.98)}\\
 \cline{2-6}
 &		\tabincell{c}{tpsv \\ a.u.}	& \tabincell{c}{6.27\\(0.86)}	&\tabincell{c}{ 6.27\\(0.86) }	& \tabincell{c}{6.09\\(1.50)}& 	\tabincell{c}{7.45\\(1.78)}\\
 \cline{2-6}
    &		\tabincell{c}{pdv \\ mm/s}	& \tabincell{c}{46.69\\(0.52) }	& \tabincell{c}{72.26\\(0.81) }	& \tabincell{c}{80.03\\(9.46)}& 	\tabincell{c}{90.64\\(13.36)}\\
 \cline{2-6}
&		\tabincell{c}{tpdv \\ a.u.}	& \tabincell{c}{22.36\\(1.30) }	& \tabincell{c}{22.36\\(1.30) }	& \tabincell{c}{22.36\\(1.30) }	& \tabincell{c}{21.18\\(0.39)}\\
 \cline{2-6}
&		 \tabincell{c}{mv \\ mm/s}	&\tabincell{c}{ 11.81\\(0.09)	}& \tabincell{c}{18.04\\(0.15) }	& \tabincell{c}{19.62\\(2.42) }	& \tabincell{c}{14.94\\(1.64)}\\
\hline
\multirow{5}{*}{\tabincell{c}{circum- \\ ferential}}&	\tabincell{c}{psv\_c \\ mm/s}	& \tabincell{c}{33.83\\(0.99)}& 	\tabincell{c}{39.88\\(0.77) }	& \tabincell{c}{35.81\\(5.75)}& \tabincell{c}{ 	32.72\\(7.40)}\\
 \cline{2-6}
&	\tabincell{c}{tpsv \\ a.u.}& 	\tabincell{c}{3.00\\(0.00)}& \tabincell{c}{	3.00\\(0.00)}& 	\tabincell{c}{4.91\\(3.94)}& 	\tabincell{c}{4.55\\(4.10)}\\
 \cline{2-6}
&	\tabincell{c}{pdv \\ mm/s}& 	\tabincell{c}{61.16\\(2.59)}& 	\tabincell{c}{42.85\\(2.07)}& 	\tabincell{c}{40.88\\(20.33)}& 	\tabincell{c}{38.67\\(18.77)}\\
 \cline{2-6}
&	\tabincell{c}{tpdv \\ a.u.} &	\tabincell{c}{22.91\\(0.67)}& 	\tabincell{c}{22.64\\(0.88)}& 	\tabincell{c}{23.09\\(1.00)}& 	\tabincell{c}{21.45\\(0.50)}\\
 \cline{2-6}
&	\tabincell{c}{mv \\ mm/s}& \tabincell{c}{	17.01\\(0.57)}& 	\tabincell{c}{18.66\\(0.39)}& 	\tabincell{c}{16.61\\(2.88)}& 	\tabincell{c}{13.88\\(1.57)}\\
\hline
\multirow{5}{*}{\tabincell{c}{longitu-\\dinal}}&	\tabincell{c}{psv \\ mm/s}	& \tabincell{c}{33.83\\(0.99)}& 	\tabincell{c}{39.88\\(0.77)}& 	\tabincell{c}{35.81\\(5.75) }	& \tabincell{c}{32.72\\(7.40)}\\
 \cline{2-6}
&	\tabincell{c}{tpsv \\ a.u.}& \tabincell{c}{ 3.00\\(0.00)}& 	\tabincell{c}{3.00\\(0.00)}& \tabincell{c}{ 	4.91\\(3.94)}& \tabincell{c}{	4.55\\(4.10)}\\
 \cline{2-6}
&	\tabincell{c}{pdv \\ mm/s}& 	\tabincell{c}{61.16\\(2.59)}& 	\tabincell{c}{42.85\\(2.07)}& \tabincell{c}{ 40.88\\(20.33)}& 	\tabincell{c}{38.67\\(18.77)}\\
 \cline{2-6}
&	\tabincell{c}{tpdv \\ a.u.}& 	\tabincell{c}{22.91\\(0.67)}& 	\tabincell{c}{22.64\\(0.88)}& 	\tabincell{c}{23.09\\(1.00)}& 	\tabincell{c}{21.45\\(0.50)}\\
 \cline{2-6}
&	\tabincell{c}{mv \\ mm/s}& \tabincell{c}{	17.01\\(0.57)}& 	\tabincell{c}{18.66\\(0.39)}& 	\tabincell{c}{16.61\\(2.88) }	& \tabincell{c}{13.88\\(1.57)}\\
\hline
\end{tabular}
\end{table}

\section{Discussion}
\label{sec:discussion}
In this work, we proposed HDL to interpolate magnitude data and to synthesise phase data of 3Dir MVMs for digital twin synthesis. Interpolation results in Section \ref{sec:interpexp} show that the proposed multi-task R2UNet has successfully interpolated magnitude images, as well as provided accurate segmentation results. Phase generation results in Section \ref{sec:phaseexp} show that phase images could be synthesised from magnitude images using a foreground-background generation scheme. Besides, velocity assessment results in Section \ref{sec:velocityexp} further prove our hypothesis that synthetic 3Dir MVM could produce an accurate myocardial motion pattern. In this section, we will discuss several problems in detail regarding 1) the choices of basic models, 2) down-sample factors for temporal interpolation, and 3) the proposed foreground-background generation scheme.
\subsection{Choice of the Basic Model for Temporal Interpolation Module}
As mentioned in Section \ref{sec:interpmethod}, we did not use a discriminator or a GAN based loss for temporal data synthesis. We hypothesised that the temporal data synthesis task should focus on generating accurate values for all pixels, and GAN based loss focuses more on the style and texture of synthesised images. In Table \ref{tab4} we presented the comparison using our UNet based model and a GAN based model for temporal data synthesis with $K\in [1,6]$ on the validating set. For the GAN based model, we used a pix2pix structure and a patch GAN discriminator for training. \textcolor{black}{The loss function of GAN based model is a combination of MSE losses (including MSE, MSEW1 and MSEW2) and the discriminator loss. The generator of this GAN based model is the same as our proposed UNet based model. As shown in Table \ref{tab4}, GAN with a discriminator loss did not improve the interpolation performance. Besides, it uses an additional generator, which introduces unnecessary parameters and slows down the optimization convergence.} 
\begin{table*}\centering
\caption{Comparison results between training temporal data synthesis module with or without the GAN loss. \textcolor{black}{With the increase of the interpolated numbers, the overall MSE difference and PSNR difference between UNet (ours) and GAN interpolated images decrease. The UNet can interpolate images with a more accurate LV position (as shown in the MSEW1 column).}}
\label{tab4}
\begin{tabular}{cccccccc}
\hline
 & & 		MSE$\times$100(Std$\times$100)&	MSEW1$\times$10(Std$\times$10)&	MSEW2$\times$100(Std$\times$100)&	PSNR(Std)&	SSIM(Std)&DICE(Std)\\
\hline
\multirow{2}{*}{$K=1$}&	Ours&	0.16(0.34)&	0.18(0.05)&	0.12(0.36)&	76.43(61.68)&	0.82(0.05)&	0.97(0.18)\\
                    &	GAN&	0.17(0.39)&	0.17(0.06)&	0.13(0.27)&	75.90(62.07)&	0.81(0.06)&	0.97(0.08)\\
\hline	 
\multirow{2}{*}{$K=2$}&	Ours&	0.15(0.21)&	0.24(0.40)&	0.15(0.21)&	76.50(58.92)&	0.81(0.07)&	0.96(0.09)\\

                    &	GAN&	0.18(0.34)&	0.21(0.31)& 0.12(0.46)&	75.42(72.86)&	0.78(0.01)&	0.93(0.04)\\
\hline
\multirow{2}{*}{$K=3$}&	Ours&	0.14(0.48)&	0.29(0.19)&	0.15(0.31)&	77.11(54.74)&	0.84(0.07)&	0.95(0.17)\\

                    &	GAN&	0.17(0.55)&	0.32(0.20)& 0.15(0.24)&	75.77(54.96)&	0.80(0.07)&	0.94(0.09)\\
\hline
\multirow{2}{*}{$K=4$}&	Ours&	0.18(0.12)&	0.37(0.10)&	0.19(0.23)&	75.75(49.25)&	0.82(0.06)&	0.94(0.18)\\

                    &	GAN&	0.33(0.08)&	0.42(0.26)&	0.23(0.45)&	73.09(41.09)&	0.79(0.01)&	0.94(0.21)\\
\hline
\multirow{2}{*}{$K=5$}&	Ours&	0.26(0.12)&	0.41(0.09)&	0.23(0.05)&	74.45(47.27)&	0.80(0.02)&	0.93(0.03)\\

                    &	GAN&	0.28(0.16)&	0.52(0.29)&	0.26(0.66)&	73.10(49  .16)&	0.73(0.04)&	0.93(0.11)\\
\hline
\multirow{2}{*}{$K=6$}&	Ours&	0.24(0.15)&	0.49(0.25)&	0.24(0.41)&	74.68(44.19)&	0.79(0.07)&	0.93(0.07)\\

                    &	GAN&	0.24(0.16)&	0.52(0.28)&	0.23(0.45)&	74.31(41.37)&	0.77(0.07)&	0.92(0.14)\\
\hline
\end{tabular}
\end{table*}

\subsection{Downsampling Factor K of the Temporal Interpolation Module}
We analyse the synthetic 3Dir MVM data from magnitude image series with different downsampling factors $K$. We used six metrics to evaluate the temporal interpolation performance, including MSE, $\omega_1$ weighted MSE (MSEW1), $\omega_2$ weighted MSE (MSEW2), PSNR, SSIM and DICE score. Here, mask $\omega_1$ and mask $\omega_2$ are illustrated in Fig. \ref{fig3}. The metrics are scaled for a better comparison.
\begin{figure}[h]
\centerline{\includegraphics[width=\columnwidth]{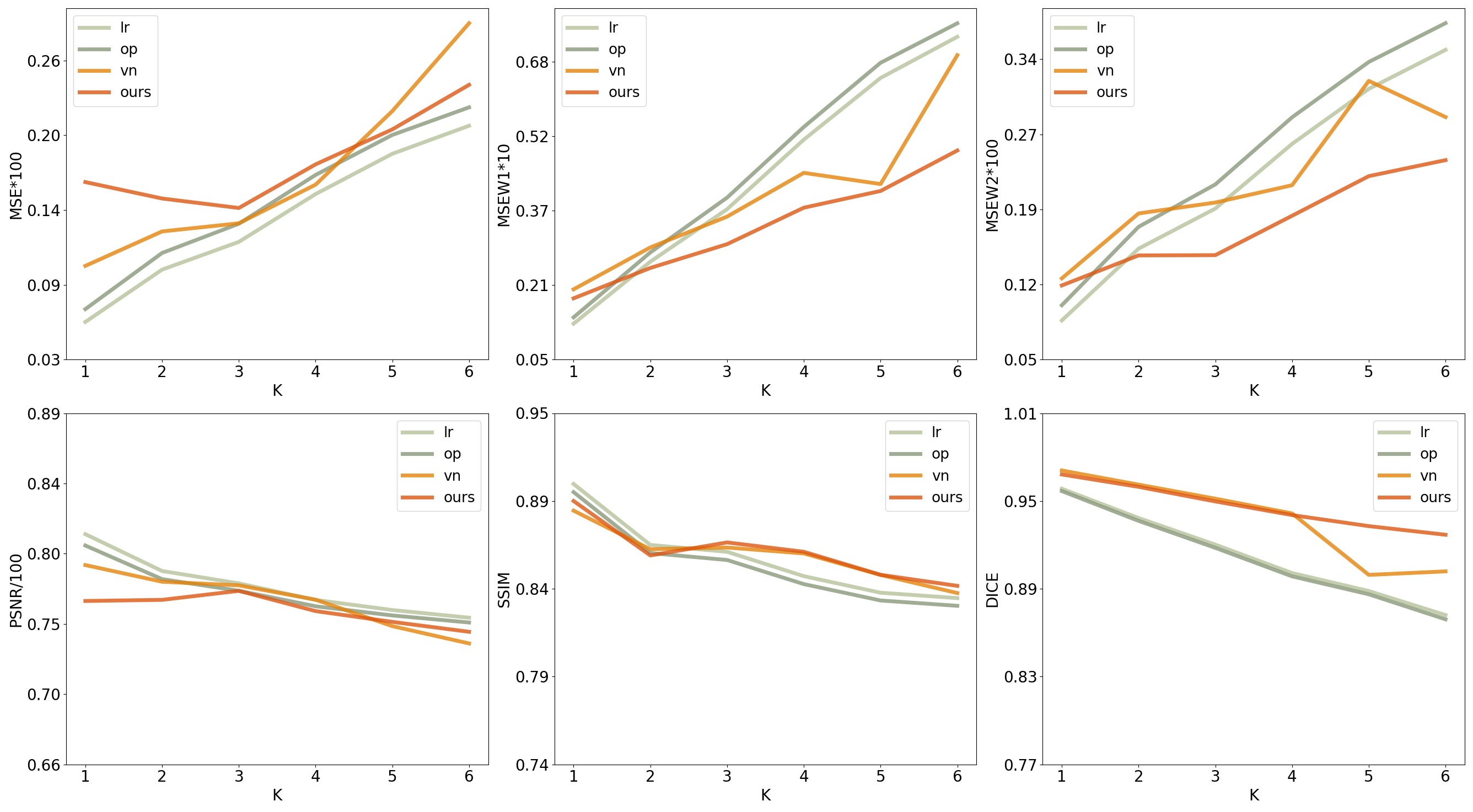}}
\caption{The synthesis performance comparison of linear interpolation (lr), optical flow (op), R2UNet (vn) and our proposed method (ours).}
\label{fig10}
\end{figure}
As is shown in Fig. \ref{fig10}, for the whole image, deep learning-based algorithms (R2UNet and ours) have a higher MSE. Because the objective functions of deep learning-based algorithms are weighted, the values on background pixels are not optimised during training. However, for weighted metrics, such as MSEW1 and MSEW2, deep learning algorithms indicate better performance. These results match our observations in Fig. \ref{fig6}, where the linear interpolated and optical flow interpolated results are more similar with ground truth in the background, while the deep learning algorithms could provide accurate ROI segmentation. The high MSE on whole images is acceptable because false predictions on the background have no effect during analysis.

When the task becomes more difficult, i.e., we discarded more images for temporal interpolation, deep learning-based algorithms could produce a more stable prediction compared to the conventional methods. This can be observed from the performance curves of the SSIM and DICE scores (Fig. \ref{fig10}). When the input magnitude image series are down-sampled 6 times, the proposed method could still produce accurate LV segmentation results with a Dice score of 0.92, which is 0.16 larger than state-of-the-art deep learning LV segmentation results \cite{wu2021fast}. 
\subsection{The Foreground-background Generating Strategy}
\begin{figure}[h]
\centerline{\includegraphics[width=\columnwidth]{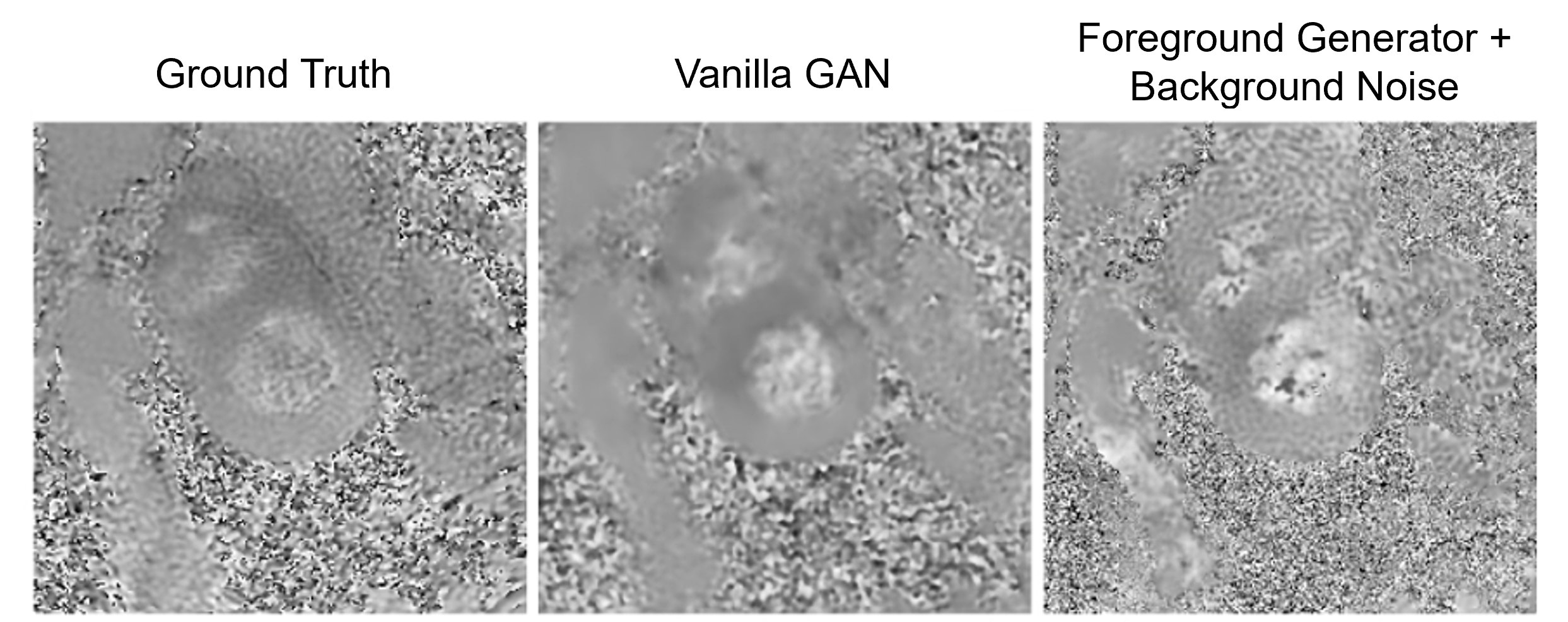}}
\caption{A visual comparison of generators with and without a foreground-background generator on phase images in the X direction. Input magnitude images are real high temporal resolution magnitude images. Our foreground generator focuses on the generation of tissue pixels and produces a sharper result compared to vanilla GAN. }
\label{fig11}
\end{figure}
In this paper, we proposed a foreground-background synthesis strategy for the phase data synthesis. We addressed the tissue pixels as foreground pixels and used a GAN based model to synthesise these pixels. In addition, we also addressed non-tissue pixels as background pixels, which were shown as random black and white pixels in phase images. We discovered that the value distribution of background pixels followed a Gaussian distribution; therefore, we used Gaussian distributions to synthesise these background pixels. 

Compared to using vanilla pix2pix, which synthesised random noise pixels and tissue pixels together, our proposed method could produce sharp predictions on the foreground pixels. This is because we optimised our network with the summation of L1 loss on whole images. The summation of L1 loss tended to smooth the synthesised images and led to failure when two parts of the image were in different distributions.
\subsection{Limitations of the current study}
There are still some limitations of the current work. Our current work has focused on the synthesis using data collected from healthy controls, and we aim to explore further studies on the conditional synthesis for patients with cardiac diseases. In the future, we will include hypertrophic cardiomyopathy (HCM) patients in this study and will test if our synthetic images could achieve the same performance on HCM/healthy classification as real images. We will also consider a multi-centre study to validate the robustness of our algorithms on varied data.

\section{Conclusion}
\label{sec:conclusion}
In this study, we have proposed an HDL framework for cardiac digital twins capable of synthesising realistic velocity maps from real-world 3Dir MVM data. Our HDL framework is composed of three modules, a U-Net based temporal interpolation module, a GAN based phase synthesis module, and a velocity assessment module on synthesised images. Our experimental results have demonstrated that synthetic data fabricated from HDL can also result in precise LV segmentation and accurate velocity assessment. The proposed HDL is the first work investigating synthetic 3Dir MVM data and provides a comprehensive and actionable pipeline for medical analysis supported by cardiac digital twins.
\bibliographystyle{IEEEtran}
\bibliography{ref.bib}

\section{Appendix}
\subsection{Network parameters}
\begin{figure*}[h]
\centerline{\includegraphics[scale=0.4]{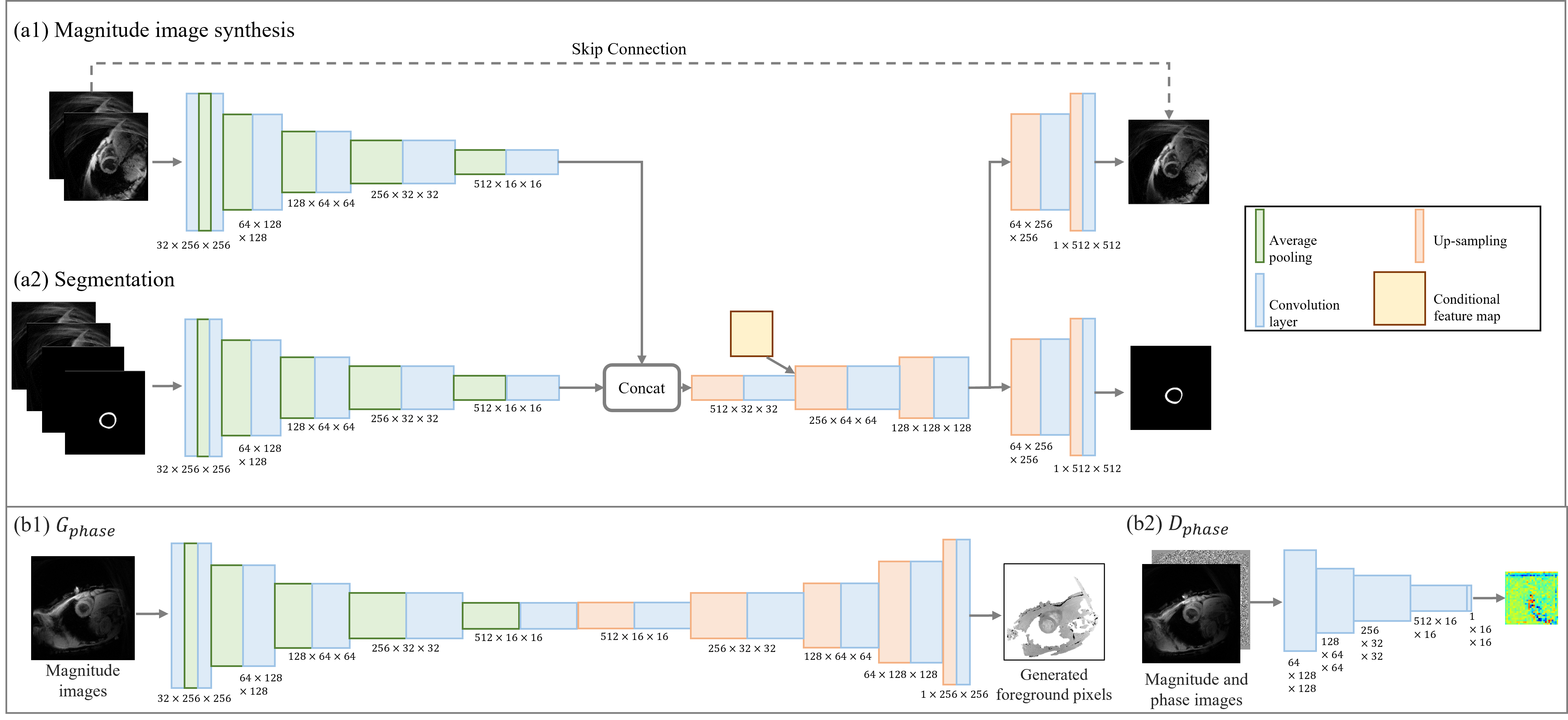}}
\caption{The detailed network parameters for temporal interpolation module (a) and phase synthesis module (b).}
\label{figp3}
\end{figure*}

The detailed network structure is shown in Fig. \ref{figp3}
\subsection{GAN and UNet comparison for paired synthesis tasks}
\begin{figure}[h]
\centerline{\includegraphics[width=\columnwidth]{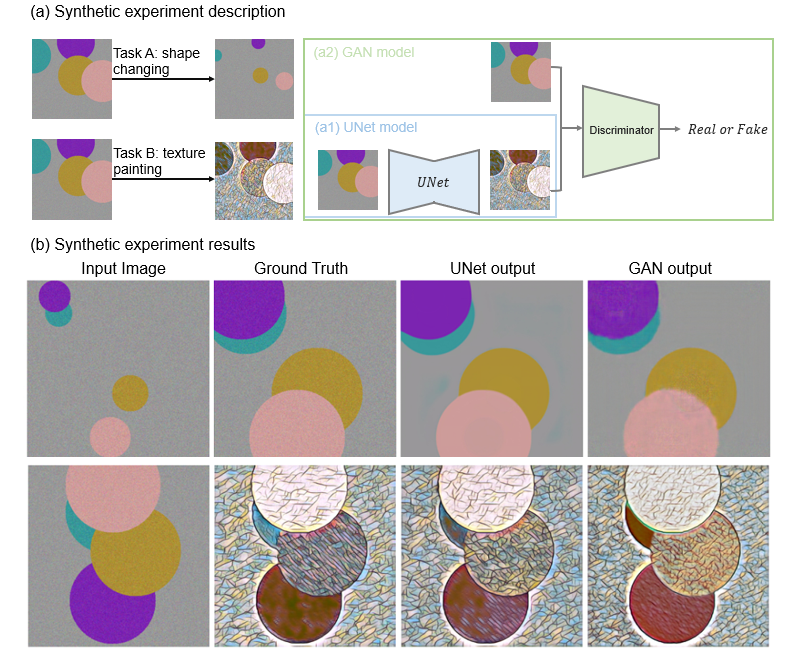}}
\caption{The design (a) and outputs (b) of the synthetic experiment. We designed two tasks for the synthetic experiment: shape-changing and texture painting. The input images are composed of four coloured circles and random noisy backgrounds.}
\label{figp1}
\end{figure}
In the literature, GAN-based algorithms are widely used for synthesis, while the UNets are regarded as not as good as GAN. However, we hypothesised that UNets work better than GANs when the style between target domain and source domain is the same, i.e., GAN-based loss, or discriminator based loss, is not recommended for frame interpolation. 

We used a synthetic experiment to prove our hypothesis. By adding circles with different values of radius and at different positions, we synthesised two datasets for two tasks: shape-changing and texture painting. The target of the two tasks is shown in Fig. \ref{figp1} (a). We used a UNet model and a GAN based model to achieve these two tasks respectively. As is shown in Fig. \ref{figp1} (b), for shape-changing, GAN loss reduces the generator performance.
\begin{figure}[h]
\centerline{\includegraphics[width=\columnwidth]{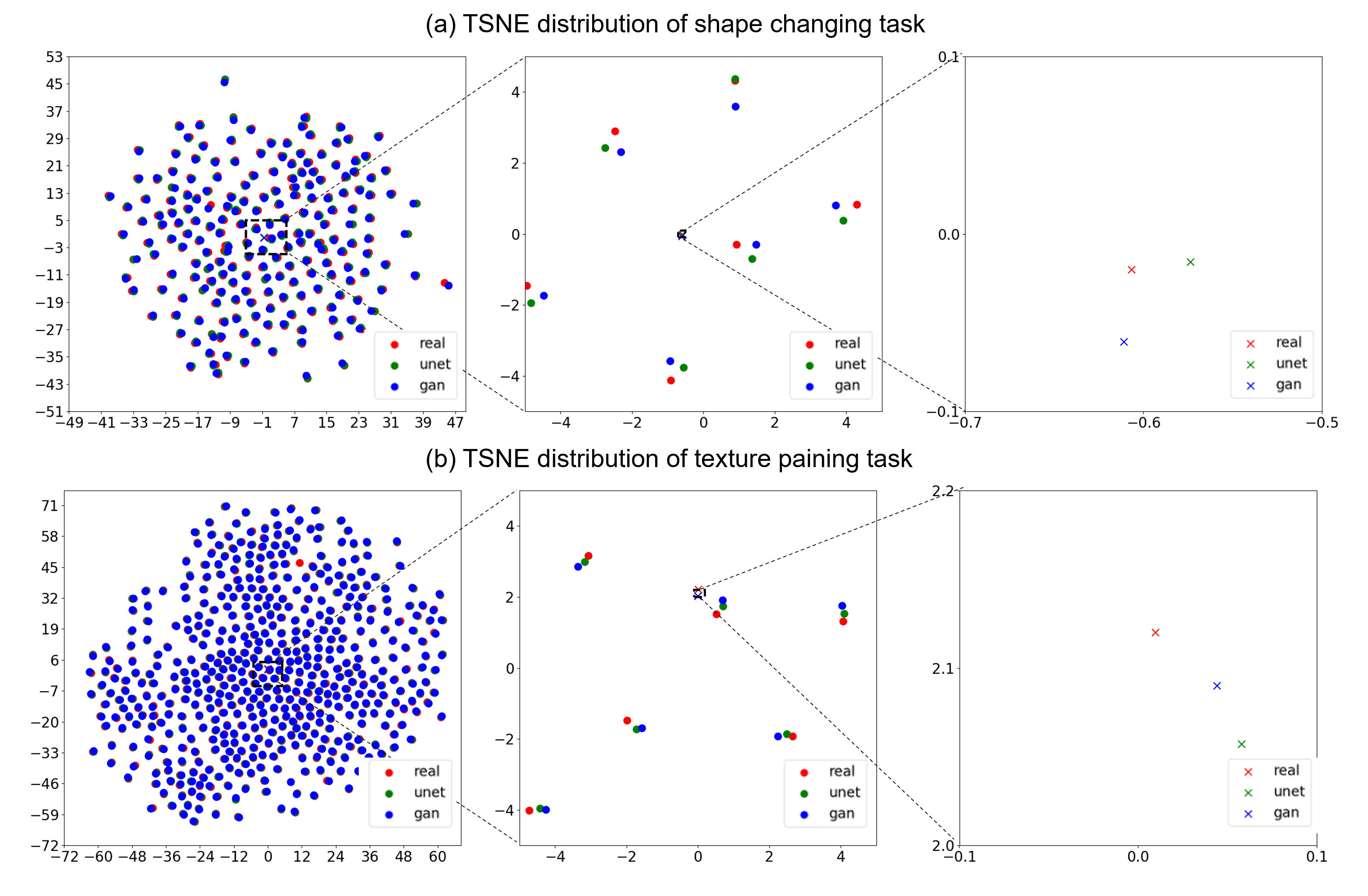}}
\caption{The data distribution of ground truth images, UNet- and GAN-synthesised images of shape changing task (a) and texture painting task (b) visualised by t-SNE. The circled point in this figure represents the individual image in this dataset, while the crosses in this figure represent the mean values of three groups. As is shown, MSE minimises the individual distance, while GAN loss minimised the group distance.}
\label{figp2}
\end{figure}
It is hard to distinguish which model performs better on texture painting tasks, so we plotted the data distribution using t-SNE in \ref{figp2}. In the middle column of \ref{figp2}, for each image, the distance between UNet synthesised images and ground truth images are closer than the distance between GAN synthesised images and ground truth images on both tasks. This is because UNet only optimises the generator with MSE loss, thus the distance between each pair of images should be closer. However, when observing the third column of Fig. \ref{figp2} (b), which is the centre of all data points from the texture painting task, the overall distribution of GAN synthesised images is closer to the ground truth distribution than that of UNet synthesised images.

\end{document}